\documentclass[]{spie}  

\usepackage{amsmath,amsfonts,amssymb}
\usepackage{graphicx}
\usepackage[colorlinks=true, allcolors=blue]{hyperref}

\usepackage{amsmath}

\usepackage{algorithm}
\usepackage{algpseudocode}
\usepackage{chngcntr}
\counterwithout{algorithm}{section}   

\usepackage{multirow}
\usepackage[table]{xcolor} 

\title{Shape Deformation Networks for Automated Aortic Valve Finite Element Meshing from 3D CT Images}

\author[a]{Linchen Qian}
\author[a]{Jiasong Chen}
\author[a]{Ruonan Gong}
\author[b]{Wei Sun}
\author[c]{Minliang Liu}
\author[a]{Liang Liang*}
\affil[a]{Department of Computer Science, University of Miami, Coral Gables, FL 33146, USA}
\affil[b]{Sutra Medical Inc, Lake Forest, CA 92630, USA}
\affil[c]{Department of Mechanical Engineering, Texas Tech University, Lubbock, TX 79409, USA}
\authorinfo{Further author information:\\
*Liang Liang: E-mail: liang@cs.miami.edu, Telephone: 1 305 284 8381}

\begin{document} 
\maketitle

\begin{abstract}
Accurate geometric modeling of the aortic valve from 3D CT images is essential for biomechanical analysis and patient-specific simulations to assess valve health or make a preoperative plan. However, it remains challenging to generate aortic valve meshes with both high-quality and consistency across different patients. Traditional approaches often produce triangular meshes with irregular topologies, which can result in poorly shaped elements and inconsistent correspondence due to inter-patient anatomical variation. In this work, we address these challenges by introducing a template-fitting pipeline with deep neural networks to generate structured quad (i.e., quadrilateral) meshes from 3D CT images to represent aortic valve geometries. By remeshing aortic valves of all patients with a common quad mesh template, we ensure a uniform mesh topology with consistent node-to-node and element-to-element correspondence across patients. This consistency enables us to simplify the learning objective of the deep neural networks, by employing a loss function with only two terms (i.e., a geometry reconstruction term and a smoothness regularization term), which is sufficient to preserve mesh smoothness and element quality. Our experiments demonstrate that the proposed approach produces high-quality aortic valve surface meshes with improved smoothness and shape quality, while requiring fewer explicit regularization terms compared to the traditional methods. These results highlight that using structured quad meshes for the template and neural network training not only ensures mesh correspondence and quality but also simplifies the training process, thus enhancing the effectiveness and efficiency of aortic valve modeling.
\end{abstract}

    
\section{INTRODUCTION}
\label{sec:intro}  

Aortic valve disease, particularly aortic stenosis, is one of the most prevalent and severe cardiac valve disorders \cite{1_lindman2016calcific}. Patient-specific computational modeling of the aortic valve has recently emerged as a powerful tool for clinical research and surgical procedural planning. High-fidelity 3D geometries of aortic valve leaflets facilitate finite element analysis to estimate biomechanical stress distributions \cite{2_structural_heart}. Such geometries can also be incorporated into fluid dynamics simulations to investigate how abnormal leaflet morphology or stiffness influences blood flow patterns in conditions such as aortic stenosis \cite{3_pase2023parametric}, which can provide valuable information for pre-operative planning of valve repair/replacement surgery \cite{4_TMI_GEOMETRY_MODELING}. Despite these advantages, generating detailed patient-specific aortic valve meshes from medical images remains a laborious and time-consuming process due to anatomical complexity and variations \cite{5_transdeformer}. Automated geometry (i.e., mesh) reconstruction approaches have been widely introduced in clinical applications. However, the effectiveness is restricted by the coarse resolution of medical images \cite{10_a_lior2023semi}, which prevents precise reconstruction of leaflet anatomy. Also, patient-specific unstructured meshes often lack consistent correspondence across different patients \cite{11_b_pak2024robust}, thus making analysis and evaluation challenging and requiring additional refinement or regularization to preserve mesh quality.

In this work, we propose a template-fitting pipeline with deep neural networks, which produces high-quality quad meshes of the aortic valve with consistent correspondence from 3D CT images, enabling simplified loss functions with only two terms while preserving mesh smoothness and element quality.

\section{RELATED WORKS}
\label{sec:related-works}

A structured quad (i.e., quadrilateral) mesh, derived from a common template, maintains consistent node-to-node and element-to-element correspondence across different patients. Such structured meshes lead to high quality and stability in further simulations. In contrast, unstructured (i.e., triangular) meshes often suffer from irregular topologies and poorly shaped elements that require additional smoothing steps to achieve acceptable mesh quality for clinical applications. 

Traditional image-based aortic valve modeling workflows involve explicit segmentation followed by surface mesh generation using algorithms such as marching cubes \cite{10_a_lior2023semi}. However, the accuracy of the segmentation-to-mesh pipelines strongly depends on the segmentation quality and the anatomical complexity of the valve \cite{8_fanwei_whole_heart}. Furthermore, the resulting meshes are unstructured and error-prone, thus requiring extensive post-processing to achieve smooth and accurate meshes for clinical applications (e.g., finite element simulations).


In contrast, template deformation-based geometry reconstruction of the aortic valve using machine learning techniques, is an ideal approach to obtain patient-specific valve meshes, as it ensures mesh correspondence across patients and thus enables consistent geometric measurements for clinical applications \cite{5_transdeformer}. Before the widespread adoption of deep learning, classical machine learning techniques were used for template initialization and non-rigid deformation \cite{6_TMI_4chamber}. For example, Liang et al. modeled aortic valve leaflet deformations from CT images using a linear coding framework with a dictionary of representative shapes \cite{7_ml_gr_liang}. More recently, deep learning models have been developed for automatic cardiac geometry reconstruction. Kong et al. proposed UNet-GCN, a hybrid framework combining a UNet backbone with a graph convolutional deformation network for whole-heart mesh generation \cite{8_fanwei_whole_heart}. Similarly, Pak et al. introduced a UNet-based model that predicts a regular-grid displacement field to deform a predefined aortic template using 80 cardiac CT scans \cite{4_TMI_GEOMETRY_MODELING}. However, these methods rely on complex combinations of multiple loss terms to simultaneously improve accuracy and preserve mesh smoothness and quality, often requiring extensive trial and error during model design and training. This challenge largely arises from the use of unstructured meshes that lack consistent correspondence across samples/patients.

In this work, our main contributions are as follows:
(1) We remesh all aortic valve geometries using a template-fitting approach during data preprocessing, producing annotated datasets with high-quality quad meshes and consistent inter-patient correspondence.
(2) We simplify the learning objective by employing only two loss terms that effectively preserve mesh smoothness and element quality, replacing unstructured triangular meshes and their complex composite loss functions.
We evaluate our framework on 3D geometry modeling from CT aortic valve images using established template-deformation networks (e.g., UNet-GCN and UNet-Disp) and assess performance via Average Point-to-Point Distance (APPD), Hausdorff Distance (HD), and Chamfer Distance across four anatomical components (aortic wall and three leaflets). In this application, our approaches with quad mesh of consistent correspondence across patients have higher accuracy with fewer loss terms applied. This resulting geometry could be used in further clinical applications such as finite element simulations.

\section{METHODS}
\subsection{Overview of Template-Based Deformation Pipeline}
\label{sec:sec-overview-pipeline}

\begin{figure} [ht]
   \begin{center}
   \begin{tabular}{c} 
   \includegraphics[width=0.8\textwidth]{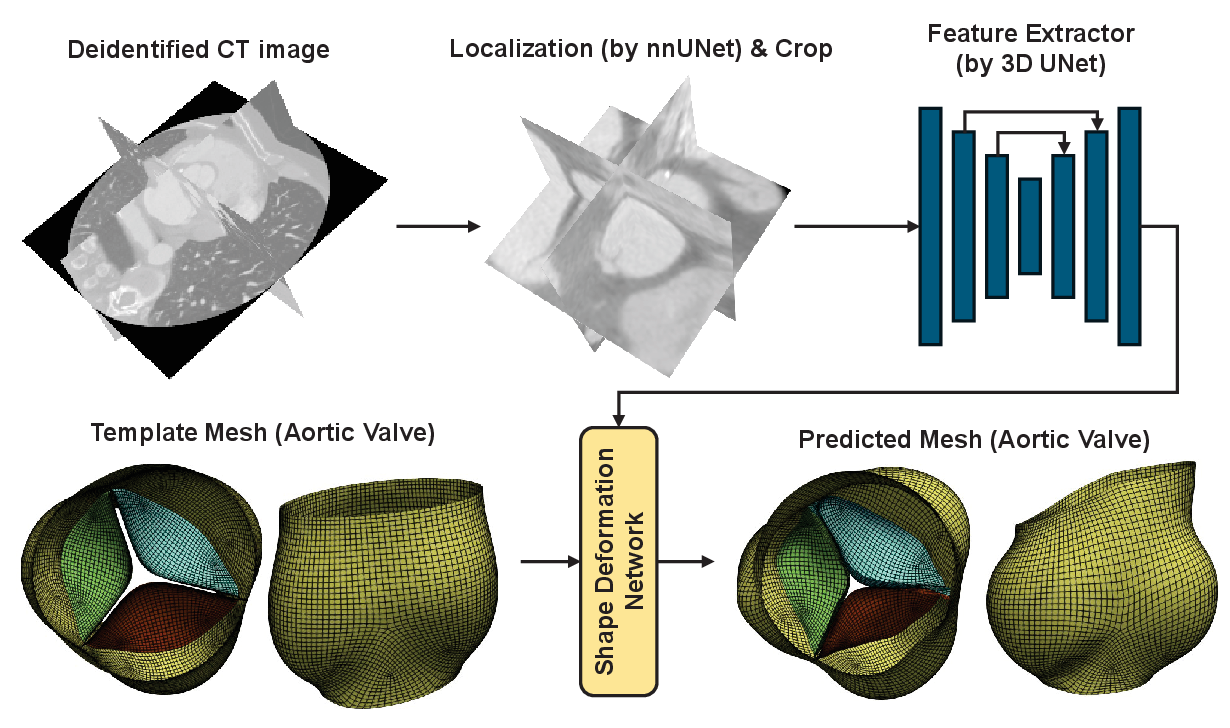}
   \end{tabular}
   \end{center}
   \caption[our-diagram] 
   { \label{fig:overview-pipeline} 
The high-level overview of the template-based deformation pipeline.}
   \end{figure}

The high-level overview of existing medical deep learning approaches for geometry modeling using template deformation is depicted in Figure. \ref{fig:overview-pipeline}. Briefly, starting from a deidentified cardiac CT image, the aortic valve region is first localized via a pre-trained deep localization network (e.g., nnUNet) and cropped according to the center of localized area \cite{9_li2022segmentation}. An image feature extractor (e.g., 3D UNet) enables compact representation of the cropped CT image \cite{4_TMI_GEOMETRY_MODELING, 8_fanwei_whole_heart} in the feature/latent space. These image features drive a shape deformation network that deforms a template aortic valve mesh to fit the aortic valve anatomy/geometry of the patient. In the existing approaches, the loss function contains a Chamfer loss term to measure the difference between a reconstructed/predicted geometry and the corresponding ground-truth geometry, and many other loss terms related to mesh smoothness and element quality, as there is no point-to-point correspondence between ground-truth geometries and reconstructed geometries. 

In our work, we follow this general pipeline and make the following innovations: we remesh all the aortic valve geometries using a novel template-fitting approach in the data preprocessing stage, and therefore data with high mesh quality and mesh correspondence are available for training and testing, which enables us to use only two loss terms in our loss function and naturally preserves mesh quality.

\subsection{Our Approach for Data Preprocessing to Obtain Ground Truth Quad Meshes}
The high-level overview of our template-fitting based remeshing algorithm is depicted in the Algorithm. 1. The algorithm is formulated as an iterative optimization that deforms a quad template mesh to fit a target polygonal surface, while enforcing mesh quality through multiple regularization terms. A target mesh is manually obtained from a 3D image of the aortic valve using 3D Slicer. The pipeline consists of affine transformation, smoothing-based relaxation, boundary constrained optimization, interior node adjustment, and final projection. 

Firstly, we apply an affine transformation to roughly align the template mesh with the target mesh under boundary nodes constraints, optimized by mean squared error. This step ensures that the key boundary features of the template mesh are initially matched to those of the target mesh. Next, Laplacian smoothing is applied to the interior nodes while boundary nodes remain fixed. Interior nodes are iteratively relaxed to reduce distortion and enhance smoothness. Together, the affine transformation and smoothing serve as initialization steps that improve mesh quality and prevent poor configurations before the main deformation. 

Then we introduce a displacement field for the entire template mesh as the main optimization variable. The optimization first focuses on boundary-constrained alignment, ensuring the template’s boundary nodes gradually and accurately match the target’s boundaries. After the boundary alignment is finalized and boundary nodes are fixed, the interior nodes continue to be optimized to improve both surface fitting and mesh quality. The objective function combines the chamfer distance between the deformed mesh and the target surface with optional weighted regularization terms. These regularizers, such as smoothness, flatness, aspect ratio, and corner angle constraints, enforce element quality and prevent collapse into irregular or crumpled shapes that could occur if chamfer distance were minimized alone. 

Finally, we project the mesh directly onto the target surface. In each iteration, nodes are mapped to their nearest nodes on the target geometry, ensuring exact surface adherence. This operation fine-tunes the deformation of the template mesh and produces a high-quality quad mesh that preserves boundary features while maintaining smoothness and structural regularity throughout the optimization.

\begin{algorithm}[H]
\caption{Template-Fitting Remeshing}
\label{alg:template_fitting}
\begin{algorithmic}[1]   
\Statex \textbf{Input:} 

    $\mathcal{X}^{0}$: quad mesh template with known topology; 
    
    $\mathcal{S}$: patient-specific surface polygon mesh; 
    
    $\mathcal{C}$: list of boundary constraints (polylines);
    
    $n_0, n_1, n_2, n_3$: number of iterations for each deformation stage
    
\Statex \textbf{Output:} 

$\mathcal{X}^\ast$: remeshed quad mesh conforming to $\mathcal{\hat{S}}$

\Statex \textbf{Process:} 

\State Initialize: $\mathcal{X}^\ast \leftarrow \mathcal{X}^{0}$
\State \textbf{Step 1: Affine Transform} on $\mathcal{X}^\ast$: $\textit{affine}(\mathcal{X}^\ast)$
\State Optimize $\mathcal{L} = \bigl\| \textit{affine}(\mathcal{X}^\ast) - \mathcal{S} \bigl\|^2$, subject to boundary constraint $\mathcal{C}$

\State \textbf{Step 2: Smooth-based Fitting} on $\mathcal{X}^\ast$
\State Optimize $\mathcal{L}_{mc}(\mathcal{X}^\ast)$, exclude boundary constraint $\mathcal{C}$

\State Re-initialize: $\mathcal{X}^{0} \leftarrow \mathcal{X}^\ast$

\State \textbf{Step 3: Boundary-Constrained Optimization}
\State Optimize $\mathcal{L} = \mathcal{L}^{chamfer}(\mathcal{X}^\ast, \mathcal{S}) + \sum_j \alpha_j \, \mathcal{L}_j(\mathcal{X}^\ast, \mathcal{S})$, where $\mathcal{L}_j$ is various regularization terms, subject to boundary constraint $\mathcal{C}$

\State \textbf{Step 4: Interior Nodes Relaxation}
\State Optimize $\mathcal{L} = \mathcal{L}^{chamfer}(\mathcal{X}^\ast, \mathcal{S}) + \sum_j \alpha_j \, \mathcal{L}_j(\mathcal{X}^\ast, \mathcal{S})$, where $\mathcal{L}_j$ is various regularization terms, exclude boundary constraint $\mathcal{C}$

\State \textbf{Step 5: Final Projection}
\State Projection on $\mathcal{X}^\ast$ to $\mathcal{S}$
\State Optimize $\mathcal{L}(\textit{projection}(\mathcal{X}^\ast), \mathcal{S})$

\State \Return $\mathcal{X}^\ast$
\end{algorithmic}
\end{algorithm}

\subsection{Our Loss Function for Training Neural Networks to Reconstruct Geometries}
\label{sec:our-loss-function-quad-mesh}
To train the networks in Figure. \ref{fig:overview-pipeline}, the loss $L$ combines a geometry term $L_\text{geom}$ and a mean curvature term $L_\text{mc}$, ensuring both mesh accuracy and quality. The geometry term $L_geom$ is the mean squared error (MSE) of mesh nodes. The mean curvature term encourages local mesh smoothness: a flat local surface has zero mean curvature. 

\begin{equation}
\label{eq:loss-function}
\mathcal{L} = \mathcal{L}_{\text{geom}} + \mathcal{L}_{\text{mc}}, 
\quad \text{where} \quad
\begin{cases}
\mathcal{L}_{\text{geom}} = \dfrac{1}{N_p} \sum_{i=1}^{N_p} 
    \bigl\| v_i^{\text{pred}} - v_i^{\text{true}} \bigr\|_2^2, \\[2ex]
\mathcal{L}_{\text{mc}} = \dfrac{1}{N_p} \sum_{i=1}^{N_p} 
    \Bigl\| \sum_{j \in \mathcal{N}(i)} (v_j - v_i) \Bigr\|_2^2 .
\end{cases}   
\end{equation}

Here, $v_i$ and $N_p$ denote the coordinate of the node-i and the number of nodes respectively. $N(i)$ means the set of neighboring nodes of $v_i$. We note that the output from our network is a quad surface mesh, and solid element mesh can be created by offsetting the surface mesh with appropriate thickness, which can then be used for finite element analysis as done in the reference paper \cite{7_ml_gr_liang}.

Following the pipeline mentioned in Section \ref{sec:sec-overview-pipeline}, we used two existing state-of-the-art deep neural networks, UNet-GCN \cite{8_fanwei_whole_heart} and UNet-Disp \cite{4_TMI_GEOMETRY_MODELING}, with our quad meshes and loss function.

\subsection{Loss Functions of the Existing approaches for Geometry Reconstruction}

Compared to our approach, the loss functions of the two existing template-based geometry reconstruction approaches \cite{8_fanwei_whole_heart, 4_TMI_GEOMETRY_MODELING} are very complex because these approaches train deep neural networks (e.g., UNet-GCN and UNet-Disp) on unstructured tri-meshes (i.e., triangular mesh) of the target (e.g., heart) surfaces. Unlike structured quad-meshes, unstructured tri-meshes have irregular connectivity and no fixed node ordering across samples, so specialized loss terms are needed to guide the deformation of the template and avoid irregularity, including Chamfer loss, normal consistency loss, edge length loss, and Laplacian loss. The Chamfer loss measures the distance between nodes of the predicted and ground-truth meshes to enforce mesh accuracy, and the normal consistency loss regulates the surface normals between the predicted and ground-truth meshes. Also, edge length loss and Laplacian loss encourages only on the predicted mesh with the uniform edge lengths and smooth surface. 

\begin{equation}
\label{eq:loss-function-existing methods}
\begin{cases}
\mathcal{L}_{\text{Chamfer}}(P,G) = 
\frac{1}{|P|}\sum_{p \in P}\min_{g \in G}\|p-g\|_2^2 
+ \frac{1}{|G|}\sum_{g \in G}\min_{p \in P}\|g-p\|_2^2, \\[2ex]
    
\mathcal{L}_{\text{normal}}(P, G) = 
\sum_{p \in P,\, g^* = \arg\min_{g \in G} \|p - g\|_2^2}
\left\| (p_1 - p) \times (p_2 - p) - n_{g^*} \right\|_2^2, \\[2ex]

\mathcal{L}_{\text{edge}}(P) =
\sum_{p \in P} \sum_{k_p \in \mathcal{N}(p)}
\left| \| p - k_p \|_2^2 - \mu_i^2 \right|, \\[2ex]

\mathcal{L}_{\text{Laplacian}}(P) =
\sum_{p \in P}
\left\|
p - 
\sum_{k_p \in \mathcal{N}(p)}
\frac{1}{|\mathcal{N}(p)|}
k_p
\right\|_2^2 .

\end{cases}   
\end{equation}

Here, $P$, $G$ are the nodes of the predicted mesh and ground-truth mesh. $p$, $g$ are some node from the node sets $P$, $G$. $|P|$, $|G|$, and $|\mathcal{N}(p)|$ are the number of nodes of the node sets $P$, $G$, $\mathcal{N}(p)$. $p_1$, $p_2$ are two nodes sharing the same face with the node $p$. Thus, the predicted surface normal at node $p$ is estimated by the cross product between two edges of the face connected to the node $p$. $n_{g^*}$ is the the ground truth surface normal at the node $g^*$. $N(p)$ means the set of neighboring nodes of the node $p$. More details for the meanings of the variables in the Equation \ref{eq:loss-function-existing methods} are explained in the reference paper \cite{8_fanwei_whole_heart}. The combination of these terms ensure the mesh accuracy and quality on the unstructured meshes. The total loss is combined depending on the training strategy. For example, In the following Section \ref{sec:dataset-and-evaluation-metrics},  we used the Equation \ref{eq:total-loss-add} as the loss function for the model UNetGCN-TriMesh-Add, and used the Equation \ref{eq:total-loss-mul} as the loss function for the model UNetGCN-TriMesh-Mul.

\begin{equation}
\label{eq:total-loss-add}
\mathcal{L}_{\text{total}} = 
\lambda_1\mathcal{L}_{\text{Chamfer}} 
+ \lambda_2\mathcal{L}_{\text{normal}} 
+ \lambda_3\mathcal{L}_{\text{edge}} 
+ \lambda_4\mathcal{L}_{\text{Laplacian}}, \\[2ex]
\end{equation}

\begin{equation}
\label{eq:total-loss-mul}
\mathcal{L}_{\text{total}} = 
\lambda_1\mathcal{L}_{\text{Chamfer}} 
* \lambda_2\mathcal{L}_{\text{normal}} 
* \lambda_3\mathcal{L}_{\text{edge}} 
* \lambda_4\mathcal{L}_{\text{Laplacian}}.
\end{equation}

\section{EXPERIMENTS}
\subsection{Dataset and Evaluation Metrics}
\label{sec:dataset-and-evaluation-metrics}

In our experiment, the dataset has 172 deidentified CT images that were obtained through a previously funded grant. Each CT volume has sufficient resolution in the aortic root region, so that the valve leaflets are clearly visible. The data were split into 121 cases for training, 19 for validation (model tuning), and 32 for held-out testing. For each case, a ground-truth aortic valve surface was obtained manually from 3D Slicer using a tool derived from a previous work \cite{7_ml_gr_liang}, and it was annotated with four components: the aortic root wall and the three aortic valve leaflets (these annotations follow the conventions described in the reference paper \cite{7_ml_gr_liang}). We applied our template-fitting remeshing algorithm to each ground-truth surface to produce a corresponding quad mesh with consistent topology. These remeshed ground truths served as the target output for our network during training and for error measurement during testing.

We compare four settings: UNetDisp-QuadMesh, UNetGCN-QuadMesh, UNetGCN-TriMesh-Add, and \linebreak UNetGCN-TriMesh-Mul. UNetDisp-QuadMesh refers to the UNet-Disp model trained with quad mesh data and our loss function as described in Section \ref{sec:our-loss-function-quad-mesh}. UNetGCN-QuadMesh refers to the UNet-GCN model trained with quad mesh data and our loss function as described in Section \ref{sec:our-loss-function-quad-mesh}. UNetGCN-TriMesh-Add refers to the UNet-GCN model trained with unstructured tri-mesh data and the loss function from an existing approach as described by Equation \ref{eq:total-loss-add}. UNetGCN-TriMesh-Mul refers to the UNet-GCN model trained with unstructured tri-mesh data and the loss function from an existing approach as described by Equation \ref{eq:total-loss-mul}. In all the four settings, the template quad mesh is the same.

\subsection{Quantitative Performance on Surface Geometry Reconstruction}
As seen in Table \ref{tab:surface-mesh-evaluation}, the models trained with our structured quad meshes and loss function (UNetDisp-QuadMesh and UNetGCN-QuadMesh) achieve substantially better accuracy than the baseline models trained with unstructured tri-mesh and very complex loss functions. For instance, considering the whole aortic valve reconstruction, the average point-to-point error (APPD) is about 2.15–2.18 mm for the UNetDisp-QuadMesh and UNetGCN-QuadMesh models, whereas the UNetGCN-TriMesh-Add/Mul models have APPD in the 2.66–2.88 mm range. This indicates roughly a 20–25\% reduction in mean error when using the quad mesh data and our loss function for training. A similar trend is observed in the Chamfer distance and Hausdorff distance. Moreover, the standard deviation of errors across the test set is consistently lower for our approach, implying that our approach not only improve accuracy but also reliability across different patients. These results confirm that enforcing a common mesh structure and simplifying the loss leads to more robust and precise predictions.

\begin{table}[ht]
\centering
\caption{Quantitative Metrics of surface geometry reconstruction for Different Models. Average Point-to-Point Distance (APPD), Chamfer Distance (CD), and Hausdorff Distance (HD) are reported (in mm) for each anatomical region (leaflets 0–2 and wall) and for the whole valve. Each entry shows the mean error and the standard deviation across the test set.}
\label{tab:surface-mesh-evaluation}
\resizebox{\textwidth}{!}{%
\setlength{\tabcolsep}{8pt}
\begin{tabular}{@{\extracolsep{\fill}}c|c|cc|cc|cc|cc@{}}
\hline
                           & model   & \multicolumn{2}{c|}{UNetDisp-QuadMesh} & \multicolumn{2}{c|}{UNetGCN-QuadMesh} & \multicolumn{2}{c|}{UNetGCN-TriMesh-Add} & \multicolumn{2}{c}{UNetGCN-TriMesh-Mul} \\
\multirow{-2}{*}{Region}   & metrics & mean                           & std    & mean                          & std   & mean                           & std     & mean               & std                \\ \hline
                           & APPD    & 1.989                          & 0.908  & {\color[HTML]{FF0000} 1.906}  & 0.986 & 2.402                          & 1.436   & 2.386              & 1.068              \\
                           & Chamfer & 0.792                          & 0.375  & 0.752                         & 0.429 & {\color[HTML]{FF0000} 0.736}   & 0.519   & 0.917              & 0.511              \\
\multirow{-3}{*}{leaflet0} & HD      & 4.490                          & 2.316  & 4.488                         & 2.393 & {\color[HTML]{FF0000} 4.389}   & 2.323   & 4.931              & 2.267              \\ \hline
                           & APPD    & {\color[HTML]{FF0000} 1.794}   & 0.921  & 1.926                         & 1.191 & 2.416                          & 1.809   & 2.388              & 1.338              \\
                           & Chamfer & {\color[HTML]{FF0000} 0.745}   & 0.456  & 0.805                         & 0.563 & 0.784                          & 0.720   & 0.923              & 0.674              \\
\multirow{-3}{*}{leaflet1} & HD      & 4.213                          & 2.279  & 4.407                         & 2.459 & {\color[HTML]{FF0000} 4.196}   & 2.448   & 4.783              & 2.779              \\ \hline
                           & APPD    & {\color[HTML]{FF0000} 1.796}   & 0.896  & 1.838                         & 1.015 & 2.298                          & 1.857   & 2.322              & 1.180              \\
                           & Chamfer & {\color[HTML]{FF0000} 0.734}   & 0.329  & 0.764                         & 0.467 & 0.753                          & 0.740   & 0.955              & 0.663              \\
\multirow{-3}{*}{leaflet2} & HD      & {\color[HTML]{FF0000} 3.962}   & 2.020  & 4.337                         & 2.391 & 4.387                          & 3.079   & 4.629              & 2.785              \\ \hline
                           & APPD    & {\color[HTML]{FF0000} 2.308}   & 1.179  & 2.342                         & 1.331 & 2.825                          & 1.217   & 3.155              & 1.291              \\
                           & Chamfer & 0.593                          & 0.434  & 0.632                         & 0.527 & {\color[HTML]{FF0000} 0.556}   & 0.414   & 0.781              & 0.519              \\
\multirow{-3}{*}{wall}     & HD      & {\color[HTML]{FF0000} 5.343}   & 3.349  & 5.659                         & 3.662 & 5.537                          & 3.147   & 5.723              & 3.075              \\ \hline
                           & APPD    & {\color[HTML]{FF0000} 2.150}   & 1.037  & 2.181                         & 1.181 & 2.665                          & 1.306   & 2.876              & 1.191              \\
                           & Chamfer & 0.595                          & 0.341  & 0.607                         & 0.401 & {\color[HTML]{FF0000} 0.552}   & 0.350   & 0.735              & 0.411              \\
\multirow{-3}{*}{Whole}    & HD      & {\color[HTML]{FF0000} 5.734}   & 3.277  & 6.013                         & 3.377 & 5.833                          & 3.153   & 5.986              & 2.924              \\ \hline
\end{tabular}
}
\end{table}

\subsection{Quantitative Performance on Anatomical Landmark Localization}
\begin{figure} [ht]
   \begin{center}
   \begin{tabular}{c} 
   \includegraphics[width=0.7\textwidth]{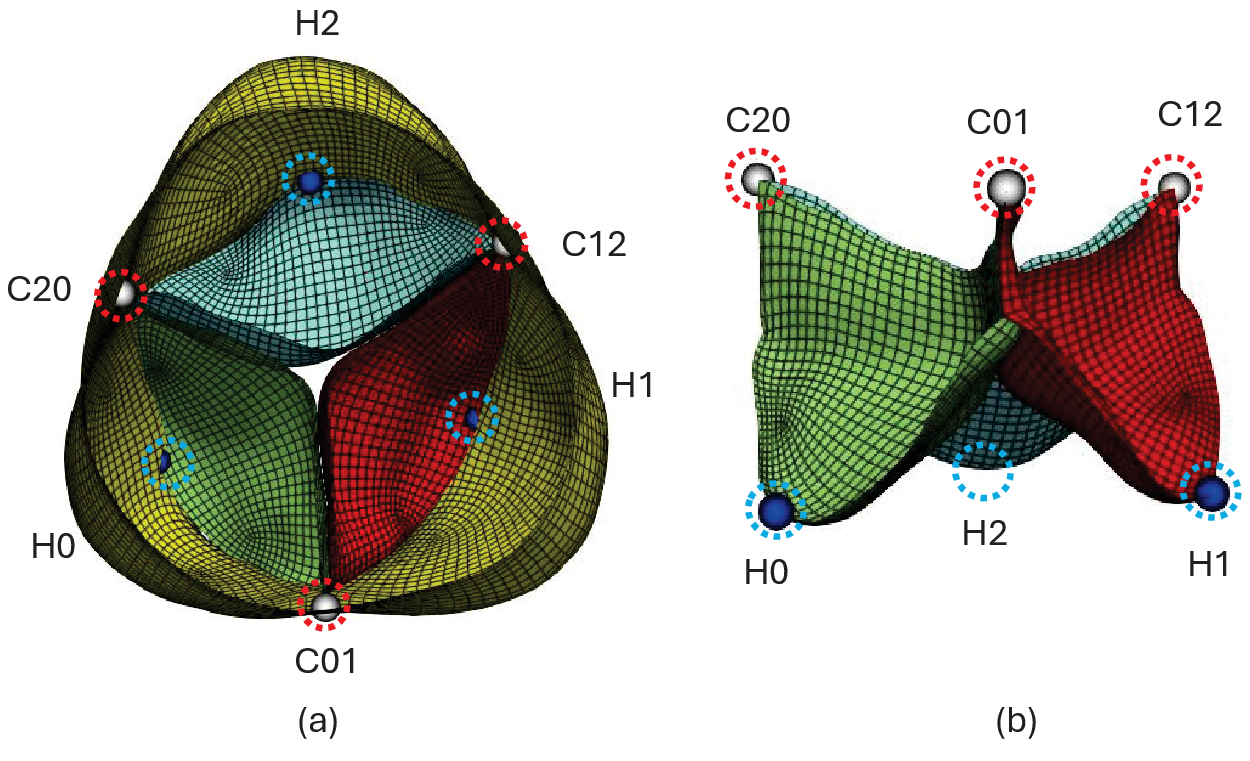}
   \end{tabular}
   \end{center}
   \caption[mesh-landmarks-diagram] 
   { \label{fig:mesh-landmarks-diagram} 
Definition of the anatomical landmarks on the mesh. Hinge nodes (H0, H1, H2) are marked with red circles and commissure nodes (C01, C12, C20) are marked with blue circles. H0 denotes the hinge node on the leaflet0 and C01 denotes thee commissure node between leaflet0 and leaflet1.}
   \end{figure}

While global surface error metrics are important, anatomical landmarks \cite{7_ml_gr_liang, 14_pat_spec_modeling_tee}, including hinge and commissure nodes, on the aortic valve are particularly critical in clinical and biomechanical contexts. For example, the commissure height is representative of the valve configuration that needs to be respected during valve-sparing procedures \cite{12_comheight_related, 13_comheight_related}. The definition of these landmarks is shown in Fig. \ref{fig:mesh-landmarks-diagram}. Thus, errors at these specific nodes could significantly affect measures like commissure height, or introduce inaccuracies in stress concentration when simulating the valve mechanics. 

The results in Table \ref{tab:error-landmark-eval} show that the models, which are trained with our structured quad meshes and loss function, achieve superior accuracy in localizing these critical landmarks. For example, the average error in hinge nodes is around 2.0–2.3 mm from the UNetDisp-QuadMesh and UNetGCN-QuadMesh models, whereas it ranges higher (often 2.5–3.2 mm) from the UNetGCN-TriMesh-Add/Mul models. By capturing commissure nodes more accurately, our approach ensures that the valve and leaflet geometry are closer to ground truth, which is important for further stress analysis. These landmark results demonstrate that our approach not only improves overall surface mesh but specifically excels at aligning crucial anatomical features of the valve.

\begin{table}[ht]
\caption{Landmark Localization Error for Hinges and Commissures. Each entry shows mean and standard deviation (in mm) for hinge nodes (H0, H1, H2) or commissure nodes (C01, C12, C20)}
\label{tab:error-landmark-eval}
\resizebox{\textwidth}{!}{%
\begin{tabular}{c|cc|cc|cc|cc}
\hline
             & \multicolumn{2}{c|}{UNetDisp-QuadMesh} & \multicolumn{2}{c|}{UNetGCN-QuadMesh} & \multicolumn{2}{c|}{UNetGCN-TriMesh-Add} & \multicolumn{2}{c}{UNetGCN-TriMesh-Mul} \\
             & mean                           & std    & mean                          & std   & mean                & std                & mean               & std                \\ \hline
Landmark-H0  & 2.007                          & 1.207  & {\color[HTML]{FF0000} 2.005}  & 1.369 & 3.171               & 2.746              & 2.652              & 1.714              \\
Landmark-H1  & {\color[HTML]{FF0000} 1.857}   & 0.978  & 2.512                         & 3.091 & 2.665               & 2.985              & 2.699              & 2.219              \\
Landmark-H2  & 2.273                          & 1.142  & {\color[HTML]{FF0000} 2.210}  & 1.459 & 2.457               & 1.590              & 2.701              & 1.462              \\
Landmark-C01 & 2.233                          & 1.666  & {\color[HTML]{FF0000} 2.139}  & 1.614 & 2.169               & 1.577              & 2.825              & 1.604              \\
Landmark-C12 & {\color[HTML]{FF0000} 2.250}   & 2.057  & 2.414                         & 2.056 & 2.585               & 2.258              & 3.318              & 3.399              \\
Landmark-C20 & {\color[HTML]{FF0000} 2.751}   & 2.325  & 2.938                         & 2.712 & 3.001               & 2.393              & 3.351              & 2.570              \\ \hline
\end{tabular}
}
\end{table}

\subsection{Qualitative Performance on Surface Geometry Reconstruction}

\begin{figure} [ht]
   \begin{center}
   \begin{tabular}{c} 
   \includegraphics[width=0.9\textwidth]{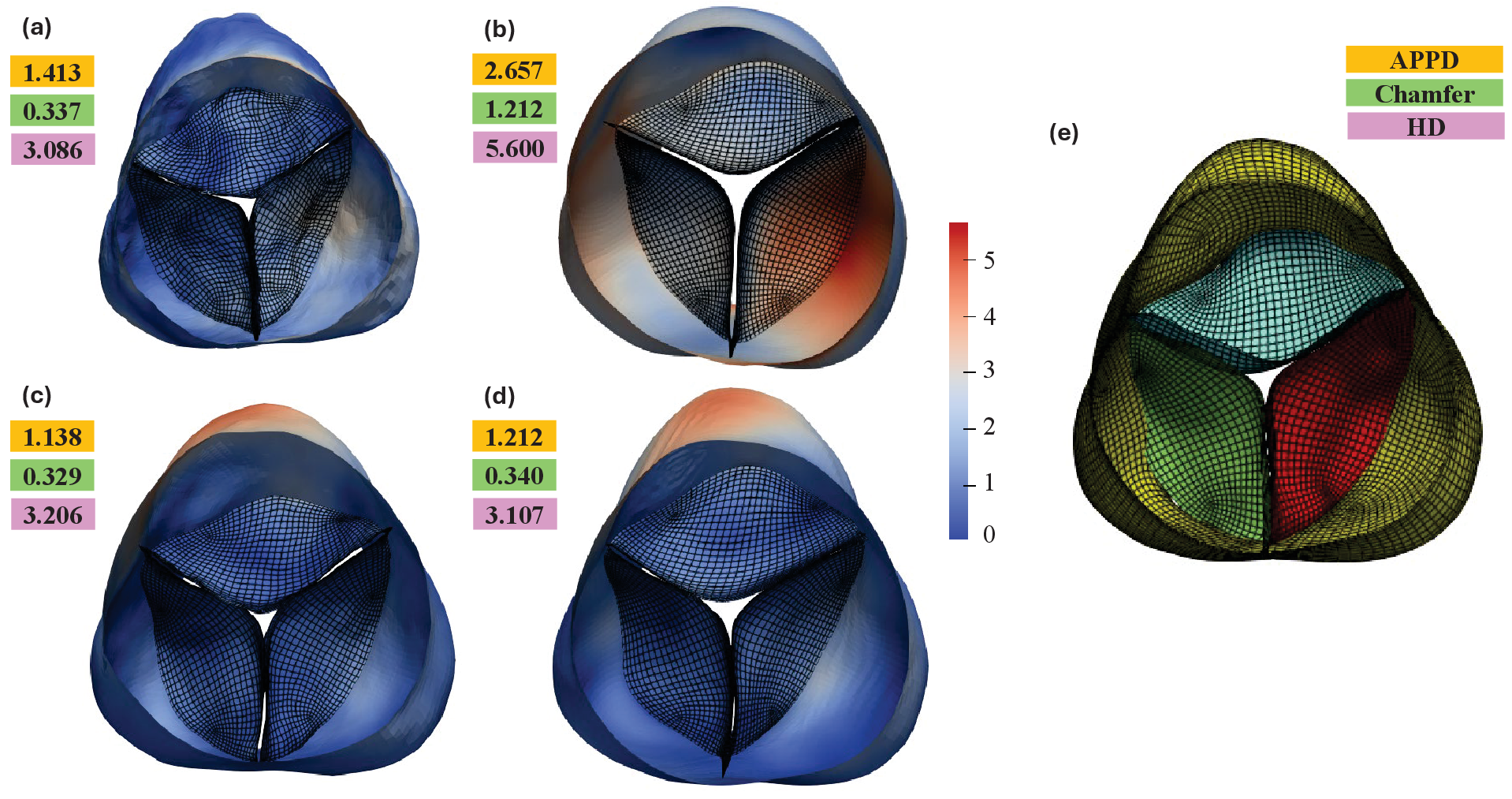}
   \end{tabular}
   \end{center}
   \caption[qual-perf-surf-mesh] 
   { \label{fig:qual-performance-surface-mesh} 
Output of the models on the specific case. (a) UNetGCN-TriMesh-Add (b) UNetGCN-TriMesh-Mul (c) UNetGCN-QuadMesh (d) UNetDisp-QuadMesh (e) Ground truth}
   \end{figure}

Figure \ref{fig:qual-performance-surface-mesh} shows representative outputs from the compared models. The baselines (UNetGCN-TriMesh-Add and UNetGCN-TriMesh-Mul) display irregularities around the leaflet edges and commissures, with distorted elements and noisy surfaces. By contrast, the models (UNetGCN-QuadMesh and UNetDisp-QuadMesh) trained by our approach produce smoother and more anatomically faithful reconstructions that closely resemble the ground truth.

These qualitative results align with the quantitative findings: in our approach, using the structured quad mesh with consistent correspondence simplifies loss function, improves accuracy, and yields regular finite element meshes.

\section{DISCUSSION}
Our findings demonstrate the impact of using a structured quad mesh template for aortic valve modeling. By remeshing all aortic valve geometries with the same template, we achieved consistent mesh correspondence across the dataset and improved overall mesh quality. By using our approach, the deep neural network models were able to accurately capture the valve shapes while maintaining smooth, well-shaped mesh elements. Moreover, this was accomplished with a simplified loss function composed of only two terms (geometry and smoothness), without the need for the many auxiliary losses that the existing approaches often require. Overall, the proposed approach ensured high-fidelity valve models with fewer training complexities, indicating a clear advantage over traditional methods based on unstructured meshing.

One major advantage of the quad mesh approach is the inherent consistency. Using a common structured mesh template means that each node and element has a fixed identity or correspondence across all patients. This consistency is critical for comparing anatomical features across patients and for training the model effectively. In our case, it ensured that comparable regions of different aortic valves (e.g., corresponding nodes on the leaflets or annulus) were aligned in the mesh representation. Such one-to-one correspondence would be difficult to guarantee with unstructured triangular meshes, which often have irregular connectivity and require complicated post-processing. Thus, the structured quad mesh simplifies downstream tasks like statistical shape analysis or point-wise error evaluation.

Another benefit of the quad mesh approach is its influence on mesh quality. Because the reconstructed geometries of the valves of different patients are deformations of the same template, the mesh topology remains consistent and mostly regular, which helps prevent irregular mesh elements. In our experiments, we observed that the deformed meshes maintained good element quality, even though we did not explicitly enforce those through separate loss terms. This indicates that the structured template with the MSE loss acted as an implicit regularizer; the smoothness term in our loss was sufficient to ensure a geometrically smooth surface, whereas a model using an unstructured mesh might need multiple additional losses to achieve comparable mesh quality. Thus, the quad mesh approach intrinsically reduces the need for many problem-specific regularization terms by providing a high-quality initialization and a constrained deformation space.

The simplified loss function (with only geometry and smoothness components) not only made training models more straightforward but also underscores why the quad mesh is more effective in this context. Fewer loss terms meant fewer hyper-parameters to tune and less risk of the model optimizing one objective at the expense of another. We found that our network could focus on aligning the template to the target valve shape and smoothing the surface, without having to employ other penalties. In contrast, prior approaches using unstructured meshes often include complex loss formulations (e.g., separate terms for mesh Laplacian smoothing, landmark correspondences, and normal consistency) precisely because the mesh lacks a consistent structure to guarantee those properties. By replacing the unstructured tri-mesh with a structured quad mesh, we essentially guarantee many of those desired properties (smoothness, correspondence, correct topology) in the data preprocessing stage. As a result, the network required minimal guidance in the form of loss terms, yet still produced accurate and smooth meshes.

In summary, using structured quad meshes for model training had a profound effect on our aortic valve geometry modeling pipeline. It allowed us to simplify the loss function while still achieving high-quality results. Future studies could build on novel approaches with quad mesh by exploring the effect of attention mechanism. In brief, our use of a quad mesh with consistent correspondence showcases a powerful strategy that combines classical mesh techniques with modern deep learning, resulting in a streamlined pipeline capable of producing accurate, simulation-ready aortic valve models.

\section{CONCLUSION}
In this work, we remeshed all the aortic valve geometries using a template-fitting approach in the data pre- processing stage to enable annotated data of high mesh quality and consistent mesh correspondence. Besides, we simplify the loss function with only two terms and preserve the mesh smoothness and element quality. The experiment demonstrates that our approach enables 3D CT aortic valve geometry modeling with state-of-the-art deep neural network models and generates geometries of mesh smoothness and element quality suitable for finite element simulation. We hope our approach can facilitate the development of clinical applications related to the aortic valve in the future.

\section*{ACKNOWLEDGMENT}
This work is supported in part by the NSF grant 2436629/2436630.

\bibliography{report} 
\bibliographystyle{spiebib} 

\end{document}